\documentclass[3p,times,twocolumn]{elsarticle}

\usepackage{ecrc}

\volume{00}

\firstpage{1}

\journalname{Nuclear Physics B Proceedings Supplement}

\runauth{D.A.~Dwyer}

\jid{nuphbp}

\jnltitlelogo{Nuclear Physics B Proceedings Supplement}

\usepackage{amssymb}

\usepackage[figuresright]{rotating}

\begin{document}

\begin{frontmatter}



\dochead{}

\title{Improved Measurement of Electron-antineutrino Disappearance at Daya Bay}

\author[label1,label2]{D.A.~Dwyer}
\author[]{on behalf of the Daya Bay Collaboration}
\address[label1]{Kellogg~Radiation~Laboratory, California~Institute~of~Technology, Pasadena, CA, USA}
\address[label2]{Lawrence~Berkeley~National~Laboratory, Berkeley, CA, USA}

\begin{abstract}
With 2.5$\times$ the previously reported exposure, the Daya Bay
experiment has improved the measurement of the neutrino mixing
parameter $\sin^{2}2\theta_{13} = 0.089\pm0.010({\rm
  stat})\pm0.005({\rm syst})$.
Reactor anti-neutrinos were produced by six 2.9 GW$_{{\rm th}}$
commercial power reactors, and measured by six 20-ton target-mass
detectors of identical design.  A total of 234,217 anti-neutrino
candidates were detected in 127 days of exposure.  An anti-neutrino
rate of $0.944\pm0.007({\rm stat})\pm0.003({\rm syst})$ was measured
by three detectors at a flux-weighted average distance of 1648 m from
the reactors, relative to two detectors at 470 m and one detector at
576 m.  Detector design and depth underground limited the background
to $5\pm0.3\%$ (far detectors) and $2\pm0.2\%$ (near detectors) of the
candidate signals.  The improved precision confirms the initial
measurement of reactor anti-neutrino disappearance, and continues to
be the most precise measurement of $\theta_{13}$.
\end{abstract}

\end{frontmatter}


\section{Introduction}
\label{sec:introduction}

The past decade has shown significant progress in the understanding of
the neutrino. In particular, oscillation of the neutrino between
electron, muon, and tau flavors has been experimentally verified.  The
amplitude of oscillation is determined by nature's choice of the
relationship between the mass and flavor eigenstates, which we
parameterize with the three mixing angles $\theta_{12}$,
$\theta_{23}$, and $\theta_{13}$.  In addition, the difference of the
squared neutrino masses, ${\Delta}m^{2}_{ji} = m^{2}_{j} - m^{2}_{i}$,
determines the frequency of flavor
oscillation~\cite{BPontecorvo:1957, BPontecorvo:1968, MNS:1962}.
As of one year ago, all of these parameters of nature had been
measured except for $\theta_{13}$.  Previous measurements of the
anti-neutrinos emitted from nuclear reactors had shown that
$\sin^{2}2\theta_{13}<0.17$ at 90\%~C.L~\cite{MApollonio:1999,
  MApollonio:2003}.  More recently, tension between measurements of
solar and reactor neutrino oscillation suggested that $\theta_{13}$
may be non-zero~\cite{GFogli:2011}.  At the end of 2011, experiments
with accelerator neutrinos~\cite{KAbe_T2K:2011, PAdamson_MINOS:2011}
and reactor anti-neutrinos~\cite{YAbe_DCHOOZ:2012} showed hints of
non-zero $\theta_{13}$ at the level of 1 to 2.5 standard deviations.

A relative measurement of reactor anti-neutrinos by detectors at two
distances was proposed to improve the sensitivity to oscillation due
to $\theta_{13}$~\cite{MikaelyanSinev:2000}.  Using this method of
multiple detectors, in Mar.~2012 the Daya Bay experiment observed a
non-zero value of $\theta_{13}$ at 5.2 standard
deviations~\cite{FPAn_PRL_DayaBay:2012}.  Soon thereafter, the RENO
experiment reported a consistent result using two
detectors~\cite{JKAhn_PRL_RENO:2012}.  The Daya Bay experiment has now
reported an improved measurement of $\theta_{13}$ using 127 days of
data, 2.5$\times$ the statistics of the initial
result~\cite{FPAn_CPC_DayaBay:2012}.

\section{The Daya Bay Experiment}
\label{sec:dayabay}

The Daya Bay experiment was designed to provide the most precise
measurement of $\theta_{13}$ of the existing and near future
experiments, with sensitivity to $\sin^{2}2\theta_{13}<0.01$ at the
90\%~C.L.  This precision was achieved through site choice, experiment
layout, and detector design.  At 17.4 GW$_{\rm th}$ the Daya Bay
reactor complex is one of the largest in the world, isotropically
emitting roughly $3\times10^{21}$ anti-neutrinos per second.  Six
anti-neutrino detectors, each with a 20-ton Gadolinium-doped liquid
scintillating target region, was the largest (4$\times$) of the
current generation of $\theta_{13}$ reactor experiments.
Anti-neutrinos were detected via inverse beta-decay (${\bar\nu}_{e} +
p \rightarrow e^{+} + n$).  The combined detection of scintillation
light from the prompt positron and the gamma rays from the delayed
capture of the neutron on Gadolinium had high efficiency ($\sim$80\%)
and low background contamination ($<$5\%).

\begin{figure}
\begin{center}
	\resizebox{\linewidth}{!}{\includegraphics{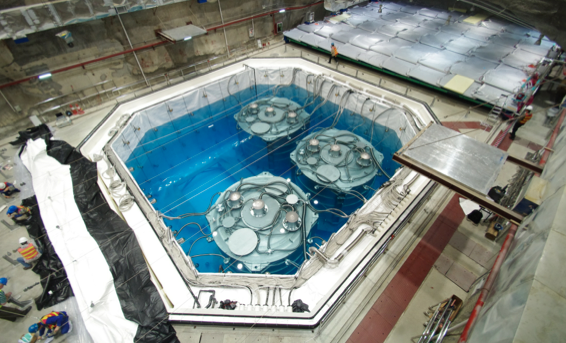}}
	\caption{Three anti-neutrino detectors (cylinders) were
          installed within the water Cherenkov muon veto in December
          of 2011.  Shortly after this photograph was taken, the
          light-tight cover (bottom left) was installed above the
          water, and the resistive plate chamber muon veto (flat
          panels in upper right) were positioned above the detectors.
          Data collection for the neutrino oscillation analysis began
          on Dec.~24, 2011.}
	\label{fig:farHall}
\end{center}
\end{figure}

The experiment layout consisted of three detector halls.  The first
hall (EH1) had two anti-neutrino detectors at a distance of 364 m from
the two Daya Bay reactor cores, EH2 had one anti-neutrino detector
roughly 500 m from the four Ling Ao reactor cores, while EH3 had three
anti-neutrino detectors 1912 m from the Daya Bay cores and 1540 m from
the Ling Ao cores.  The detectors in the near halls provided a
measurement of the mostly unoscillated total reactor flux, while the
far hall was situated near the predicted oscillation maximum.  The
experimental halls were located underground beneath adjacent
mountains. The resulting cosmic ray shielding limited cosmogenic
backgrounds to $\sim$0.2\% of the measured anti-neutrino signal.
Additionally, detectors were placed within active water Cherenkov veto
detectors.  The water veto detectors reduced cosmogenic neutrons to a
negligible background.  Figure~\ref{fig:farHall} shows the three
anti-neutrino detectors during installation in the far experimental
hall.

These experiment design choices resulted in a nearly background-free
measurement of 200 (2000) anti-neutrino interactions per day for the
far (near) detectors.  The use of multiple detectors of identical
design allowed for 0.2\% uncertainty in the relative detector
efficiency.  The hall locations were optimized to limit the impact of
uncertainties in relative neutrino flux for each reactor core.  A
complete description of the Daya Bay detectors, experiment layout, and
initial results are given in~\cite{DayaBay_Proposal:2006,
  FPAn_NIM_DayaBay:2012,FPAn_PRL_DayaBay:2012}.

\section{Neutrino Oscillation Analysis}
\label{sec:oscillation}

Using 127 days of data taken between Dec.~24,~2011 and May~11,~2012, a
total of 234,217 inverse beta-decay anti-neutrino candidates were
measured by the six detectors (EH1: 69121, 69714; EH2: 66473; EH3:
9788, 9669, 9452).  The analysis approach closely follows that
reported in~\cite{FPAn_PRL_DayaBay:2012}, except for the increase in
livetime.  The muon rates vary for each hall due to the different
mountain overburdens.  The resulting variation in muon veto times
leads to the largest difference in relative efficiencies between
detectors in separate halls (EH1: 0.8015, 0.7986; EH2: 0.8364; EH3:
0.9555, 0.9552, 0.9547), although the uncertainty in this veto time is
negligible ($<$0.01\%).  The uncertainty in relative detector
efficiencies is dominated by the uncertainty introduced by the 6-12
MeV energy selection of the delayed neutron capture (0.12\%),
described in detail in~\cite{FPAn_NIM_DayaBay:2012}.

The largest background in the anti-neutrino candidates are two
uncorrelated background radiation interactions that accidentally
satisfy the energy and time correlation for inverse beta-decay
anti-neutrino selection.  Such events, called {\em accidentals},
account for 4.0$\pm$0.05\% (1.5$\pm$0.02\%) of the far (near)
candidates.  The high statistics of uncorrelated signals allow for the
high-precision estimation of this background.  The uncertainty in
background is dominated by two sources of correlated signals.  The
cosmogenic $\beta$-n isotope $^9$Li contributes 0.3$\pm$0.2\%
(0.4$\pm$0.2\%) of the far (near) candidates.  In addition, the three
0.5~Hz $^{241}$Am-$^{13}$C neutron calibration sources present above
the lid of each detector introduced a background.  The prompt gamma
signal from neutron inelastic scattering on iron, followed by the
delayed gamma rays produced by capture on stainless steel are
estimated to be 0.3$\pm$0.3\% of the far detector candidates.  Other
correlated backgrounds due to energetic cosmogenic neutrons (i.e.\ {\em
  fast neutrons}) and ($\alpha$,n) nuclear interactions are negligible
in comparison.

\begin{figure}
\begin{center}
	\resizebox{\linewidth}{!}{\includegraphics{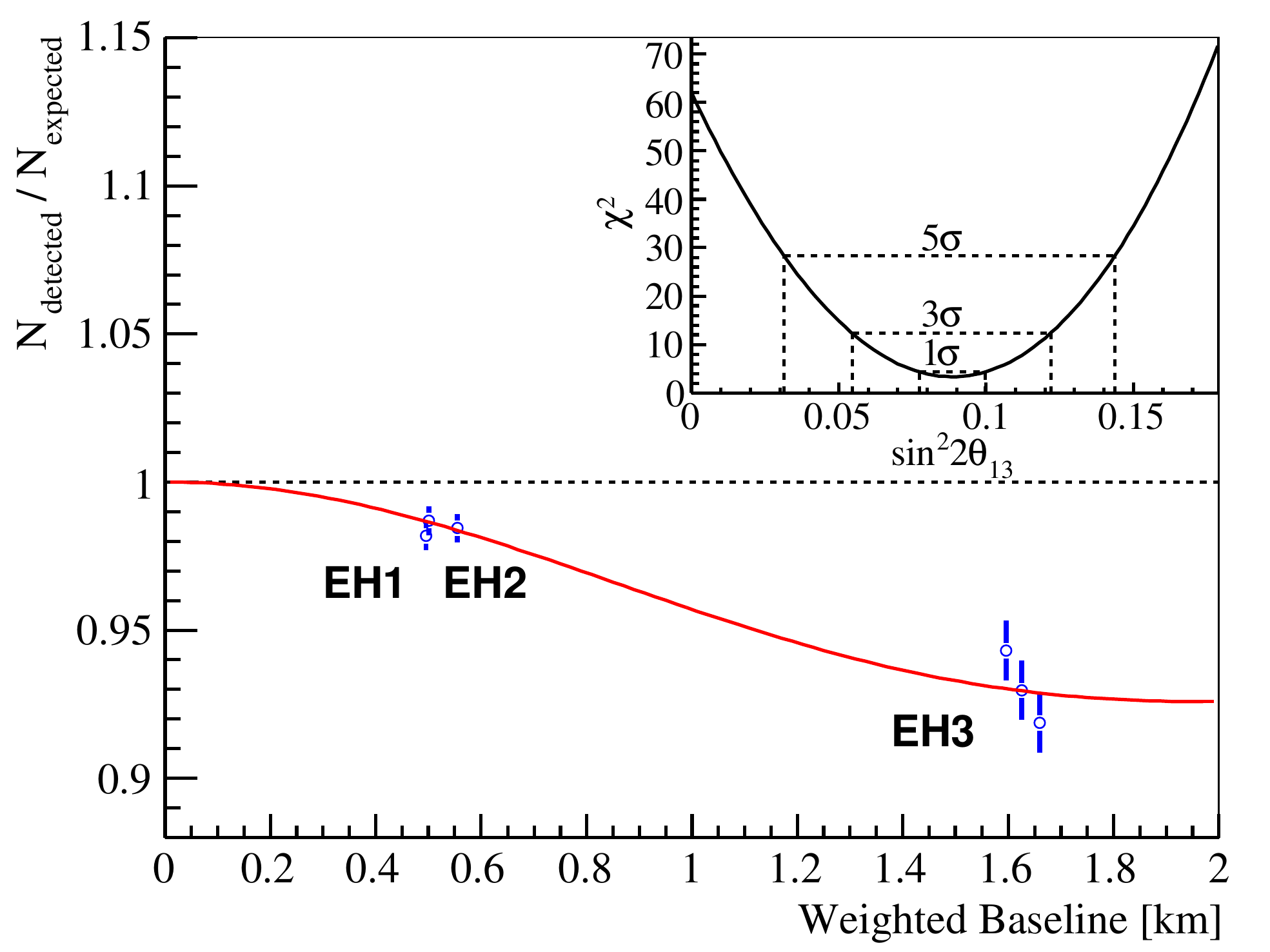}}
	\caption{A comparison of the relative detected anti-neutrino
          rates in the three near and three far detectors.  In absence
          of oscillation, all points should lie on a horizontal line.
          The curve shows the relative anti-neutrino rates due to the
          best estimate of $\sin^{2}2\theta_{13}=0.089$ in a
          three-flavor oscillation model.}
	\label{fig:rateCompare}
\end{center}
\end{figure}

After correcting the candidate counts for total livetime, veto
efficiency, and background contributions, the resulting anti-neutrino
rates in each detector were EH1: 662.47$\pm$3.00, 670.87$\pm$3.01;
EH2: 613.53$\pm$2.69; EH3: 77.57$\pm$0.85, 76.62$\pm$0.85,
74.97$\pm$0.84 per day.  The rates for detectors in the same halls are
consistent within statistical uncertainty taking into account the
small differences in distance from reactors.  The near detector
anti-neutrino rates were used to predict the expected rates at the far
detectors assuming no oscillation.  The far detectors found
0.944$\pm$0.007(stat)$\pm$0.003(syst) of the expected flux relative to
the near sites.  An analysis of the relative anti-neutrino rates of
the six detectors in a standard three-flavor oscillation model found
$\sin^{2}2\theta_{13} = 0.089\pm0.010({\rm stat})\pm0.005({\rm syst})$
[Fig.~\ref{fig:rateCompare}].

\section{Conclusion}
\label{sec:conclusion}

The addition of 77 days to the previous 50 days of Daya Bay detector
exposure has confirmed the previous observation of reactor
anti-neutrino disappearance, and improved the precision of the
estimate of the neutrino mixing angle $\theta_{13}$.  The combination
of experiment location, layout, and detector design has yielded the
highest signal rate, lowest background rate, and consequently most
precise $\theta_{13}$ measurement of the existing oscillation
experiments.  The statistical uncertainty is still twice as large as
the systematic uncertainty, so the collection of additional data will
continue to improve the precision.  The Daya Bay experiment also has
the potential to estimate the squared-mass difference
${\Delta}m^{2}_{31}$ via distortion of the positron energy
spectrum.\footnote{To be precise, electron neutrino disappearance
  experiments are sensitive to a combination of ${\Delta}m^{2}_{31}$
  and ${\Delta}m^{2}_{32}$.  The resulting {\em effective}
  squared-mass difference is commonly referred to as
  ${\Delta}m^{2}_{ee}$.}  The existing data also provides for a
high-statistics measurement of the absolute reactor anti-neutrino
flux.  Measurement of a non-zero value for $\theta_{13}$ allows for
future experiments to probe the possibility of charge-parity (CP)
violation in neutrinos.

\section*{Acknowledgments}

The Daya Bay experiment is supported in part by the Ministry of
Science and Technology of China, the United States Department of
Energy, the Chinese Academy of Sciences, the National Natural Science
Foundation of China, the Guangdong provincial government, the Shenzhen
municipal government, the China Guangdong Nuclear Power Group,
Shanghai Laboratory for Particle Physics and Cosmology, the Research
Grants Council of the Hong Kong Special Administrative Region of
China, University Development Fund of The University of Hong Kong, the
MOE program for Research of Excellence at National Taiwan University,
National Chiao-Tung University, and NSC fund support from Taiwan, the
U.S. National Science Foundation, the Alfred~P.~Sloan Foundation, the
Ministry of Education, Youth and Sports of the Czech Republic, the
Czech Science Foundation, and the Joint Institute of Nuclear Research
in Dubna, Russia. We thank Yellow River Engineering Consulting Co.,
Ltd.\ and China railway 15th Bureau Group Co., Ltd.\ for building the
underground laboratory. We are grateful for the ongoing cooperation
from the China Guangdong Nuclear Power Group and China Light~\&~Power
Company.







\end{document}